\documentclass[aip,amsmath,amssymb,reprint, twocolumn]{revtex4-1}

\usepackage{times,amsmath}
\usepackage{epsfig}
\usepackage{color}
\usepackage{longtable}
\usepackage{graphicx}
\usepackage{dcolumn}
\usepackage{bm}
\usepackage{bookmark}
\usepackage{tabularx}
\usepackage{hyperref}
\usepackage{multirow}
\usepackage{array,mathtools,amssymb,booktabs,makecell}
\hypersetup{colorlinks=true, citecolor=blue, filecolor=blue, linkcolor=blue, urlcolor=blue}

\usepackage{proofread}
\usepackage[flushleft]{threeparttable}
\begin{document}

\title{The role of surface spin polarization on ceria-supported Pt nanoparticles}

\author{Byungkyun Kang}
\affiliation{Department of Physics and Astronomy, University of Nevada, Las Vegas, NV 89154, USA}

\author{Joshua L.Vincent}
\affiliation{School for Engineering of Matter, Transport, and Energy, Arizona State University, Tempe, AZ 85281, USA}

\author{Peter A. Crozier}
\affiliation{School for Engineering of Matter, Transport, and Energy, Arizona State University, Tempe, AZ 85281, USA}

\author{Qiang Zhu}
\affiliation{Department of Physics and Astronomy, University of Nevada, Las Vegas, NV 89154, USA}

\date{\today}

\begin{abstract}
In this work, we employ  first-principles simulations to investigate the spin polarization of CeO$_2$-(111) surface and its impact on  interactions between a ceria support and Pt nanoparticles. For the first time, we report that the CeO$_2$-(111) surface exhibits a robust surface spin polarization due to the internal charge transfer between atomic Ce and O layers. In turn, it can lower the surface oxygen vacancy formation energy and enhance the oxide reducibility. We show that the inclusion of spin polarization can therefore  significantly reduce the major activation barrier in the proposed reaction pathway of CO oxidation on ceria-supported Pt nanoparticles. For metal-support interactions, surface spin polarization enhances the bonding between Pt nanoparticle and ceria surface oxygen, while CO adsorption on Pt nanoparticles weakens the interfacial interaction regardless of spin polarization.
\end{abstract}


\vskip 300 pt

\maketitle

\section{INTRODUCTION}
In heterogeneous catalysis,  metal nanoparticles supported on reducible oxide structures provide unique interfacial interactions which lead to the formation of active sites at the three-phase boundaries which control activity and selectivity for oxidation reactions~\cite{peter_nchem2011,yusuke_nchem2011}. Many high-performance heterogeneous catalysts have been developed which exploit the metal/oxide interface properties for reactions such as CO oxidation at room and low temperature~\cite{jian_jacs2019,shaofei_acscatal2019,huiyuan_jacs2015,yarong_acsapplmaterinter2020}. The ability of reducible oxides to donate lattice oxygen during a reaction (the so-called Mars van Krevellen mechanism) was first described nearly 70 years ago and is now an accepted kinetic pathway for many reactions~\cite{mars_ces1954}. The atomic level details of how such a mechanism operates especially in the presence of metallic nanoparticles remains poorly understood. Tauster was one of the first to recognize the unique properties of the metal-support interaction (MSI) with his observation of metal nanoparticles being encapsulated with thin overlayers under strong reducing condition~\cite{tauster_jacs1978,tauster_science1981,tauster_acr1987}. The strong MSI has been realized in redox transition metal oxides~\cite{tauster_jacs1978,tauster_science1981,tauster_acr1987,hailian_sciadv2017} and even a relatively redox-inert alkaline metal oxide MgO~\cite{wang_nature2021}. The MSI is also associated with occurrence of charge transfer at the interface~\cite{gianfranco_csr2018}. Campbell proposed the electronic MSI mechanism, in which charge transfer at the metal/oxide interface modulates chemical activity of the supported catalyst \cite{charles_natchem2012}. The electronic MSI gives rise to a rearrangement of electrons within the interface and enhances the rate of surface oxygen vacancy creation/annihilation leading to an enhanced catalytic activity~\cite{gianfranco_pccp2013}. 

To manipulate the role of the interfacial interactions on a catalytic reaction, the nanoparticle size can be varied to change the contact area between the  metal and support interface~\cite{insoo_acscatal}. For Au/Co$_{3}$O$_{4}$ and Au/Fe$_{2}$O$_{3}$, the gold nanoparticles with a diameter smaller than 5 nm show high reaction activity~\cite{masatake_catatoday1997} and Pt particles of 2-3 nm were found to be more  active for CO oxidation~\cite{alexey_applcatab2012}. In the limiting case, the metal nanoparticles may be reduced all the way down to a single atom, giving rise to the so-called single atom catalysts~\cite{liu2017catalysis}.
Another approach to manipulate catalytic activity is by tuning the interfacial interaction directly. This has been realized by varying the support type, particles shape and size. The aforementioned nanostructuring creates new metal-support interaction resulting in the metal nanoparticle properties being substantially different from their bulk counterparts~\cite{insoo_acscatal}. Doping oxides by heteroatoms can modify the electronic structure. They enhance surface oxygen reducibility by tuning the interfacial interaction~\cite{antonio_acscatal2017}. The increased oxide reducibility linked to surface oxygen vacancy formation energy is an indispensable element in oxidation reactions based on the Mars-van Krevelen mechanism~\cite{antonio_acscatal2017,yang_jacs2013,huan_ass2014,kuo_acscat2012,daniel_ang2011,hyun_jacs2011}. In combination with the smaller nanoparticles, nano-structured oxide supports further lower the surface oxygen vacancy formation energy and gives rise to a reverse spillover of the surface oxygen on to the Pt nanoparticle~\cite{georgi_nature2011}. These  works highlight the importance of undertaking fundamental studies to elucidate the complex interfacial interaction underlying catalytic functionality.

There remain a paucity of information about the atomic structure and structural dynamics of an active metal-support interface performing catalysis. Recently, there has been an emerging paradigm that has roots in surface science~\cite{somorjai1991flexible,imbihl1995oscillatory,ertl2008reactions} and chemistry~\cite{cotton1975fluxionality} for  understanding catalytically active sites in terms of dynamic, meta-stable, or so-called fluxional species from both computation~\cite{zhang2020ensembles,sun2018metastable,zhai2017fluxionality} and  experimental work~\cite{lawrence2021atomic,li2021dynamic,vincent2021fluxional}. This has raised many  questions regarding both the fundamental structure of active sites and the atomic scale evolution of the interface, metal particle, and adjacent oxide surface during catalysis. For example, what is the dynamic nature of the metal-support interface? The adhesion between the metal particle and support may weaken significantly since bridging oxygen are responsible for anchoring the metal to the support. How does the metal-support interface change in the presence of reactant adsorbates and reaction intermediates? How do such structural changes impact activation of intermediates and facilitate the bond breaking and formation along the reaction pathways? What happens during the rate limiting step and where is the likely site for CO$_{2}$ formation and desorption. In order to deepen our understanding of the factors affecting catalysis and to develop strategies for improved catalyst design, it is essential to elucidate and describe the structural evolution that occurs at the atomic level during simultaneous catalytic turnover.

In this work, we investigated the interfacial interaction of ceria-supported Pt nanoparticles by density functional theory (DFT) simulations. We found that spins are polarized on the CeO$_{2}$-(111) surface through charge transfer from surface oxygen to cerium, which can significantly lower the oxygen vacancy formation energy and alter the interfacial interaction between Pt nanoparticle and ceria support. In turn, the activated surface plays an essential role in lowering the major activation barrier in the proposed reaction pathway of CO oxidation on the ceria-supported Pt nanoparticles. Using the existence of a robust surface spin polarization, we propose a theoretical description of the recently observed controversial room temperature magnetism in ceria nanostructures~\cite{karl_physrep2018}.
\section{Computational Methods}
We used the slab model with five O-Ce-O layers to simulate the CeO$_{2}$-(111) surface with a 6 $\times$ 6 reconstruction of the primitive unitcell. Experimental lattice constants ($a$=$b$=5.410 $\textrm{\AA}$)~\cite{lc_exp} of ceria with 15 $\textrm{\AA}$ vacuum along the $z$ direction have been used throughout. To make Pt(111) interface with CeO$_{2}$-(111), a hexagonal Pt$_{19}$ single layer (SL) was built. As initial structure for geometry optimization, we have located the center of Pt$_{19}$ SL 2.5 $\textrm{\AA}$ above oxygen on the CeO$_{2}$-(111) surface, while Pt$_{19}$ SL [100] and CeO$_{2}$ [110] are coincided. This model was relaxed at the level of DFT by using the CP2K package~\cite{cp2k}. In CP2K, the wavefunction was expanded in the double $\zeta$ valence plus polarization and plane-wave basis sets with an energy cutoff of 400 Ry. We used Geodecker\text{-}Teter\text{-}Hutter pseudopotentials~\cite{cp2k_potential} based on the GGA~\cite{pbe} functional. The van der Waals interaction was also considered according to the DFT-D3 scheme~\cite{stefan_jcp2010}. For all spin polarized calculations, half spin moments on Pt and Ce atoms were given as the initial guess. To account for the strongly correlated nature of Ce's 4$f$ electrons, we employed the DFT+$U$ method with $U$ = 7 eV, as suggested by the previous studies ~\cite{he2018size,yang_ncomm2014}. To check the effect of $U$, we also repeated some representative calculations using $U$ = 4 eV. The climbing image-nudged elastic band (CI-NEB) method ~\cite{cineb} was used to simulate the activation barriers for the proposed reaction pathway of CO oxidation on the ceria-supported Pt nanoparticles. To find surface spin polarization effect on the lattice dynamic reconfiguration, we performed the \textit{ab-initio} molecular dynamic (AIMD) simulation at 300 K with 1 fs time step using Nose-Hover thermostat. To prevent high computational cost, only three layers of O-Ce-O were considered in the AIMD simulation. In all simulations, the atoms on the lowest O-Ce-O layer were kept fixed.

From the aforementioned DFT simulations, we obtained the adsorption energy using the following equation:
\begin{equation}
    E_{\textrm{ads}}=E_\textrm{(slab\_adsorbate)}-E_\textrm{slab}-E_\textrm{adsorbate}
\end{equation}
where $E_\mathrm{\text{(slab\_adsorbate)}}$ is the energy of optimized slab model with adsorbates, $E_\mathrm{\text{slab}}$ is the energy of pure surface slab, and $E_\mathrm{\text{adsorbate}}$ is the energy of adsorbate, respectively.

Oxygen vacancy formation energy is
\begin{equation}
    E_{\textrm{Vo}}^{\textrm{F}}=E_\textrm{(slab\_Vo)}-E_\textrm{slab}+1/2E_{\textrm{O$_{2}$}}
\end{equation}
where $E_\mathrm{\text{O}_{2}}$ is the total energy of the ground (triplet) state of oxygen molecule in the gas phase.

Surface energy of slab is defined by
\begin{equation}
    E_{\textrm{surf}}=(E_\textrm{slab}-nE_\textrm{bulk})/(2S)
\end{equation}
where $n$ is the number of formula unit in the slab, and $S$ is the surface area.

\section{RESULT AND DISCUSSIONS}

\begin{figure}[ht]
\centering
\includegraphics[width=0.46 \textwidth]{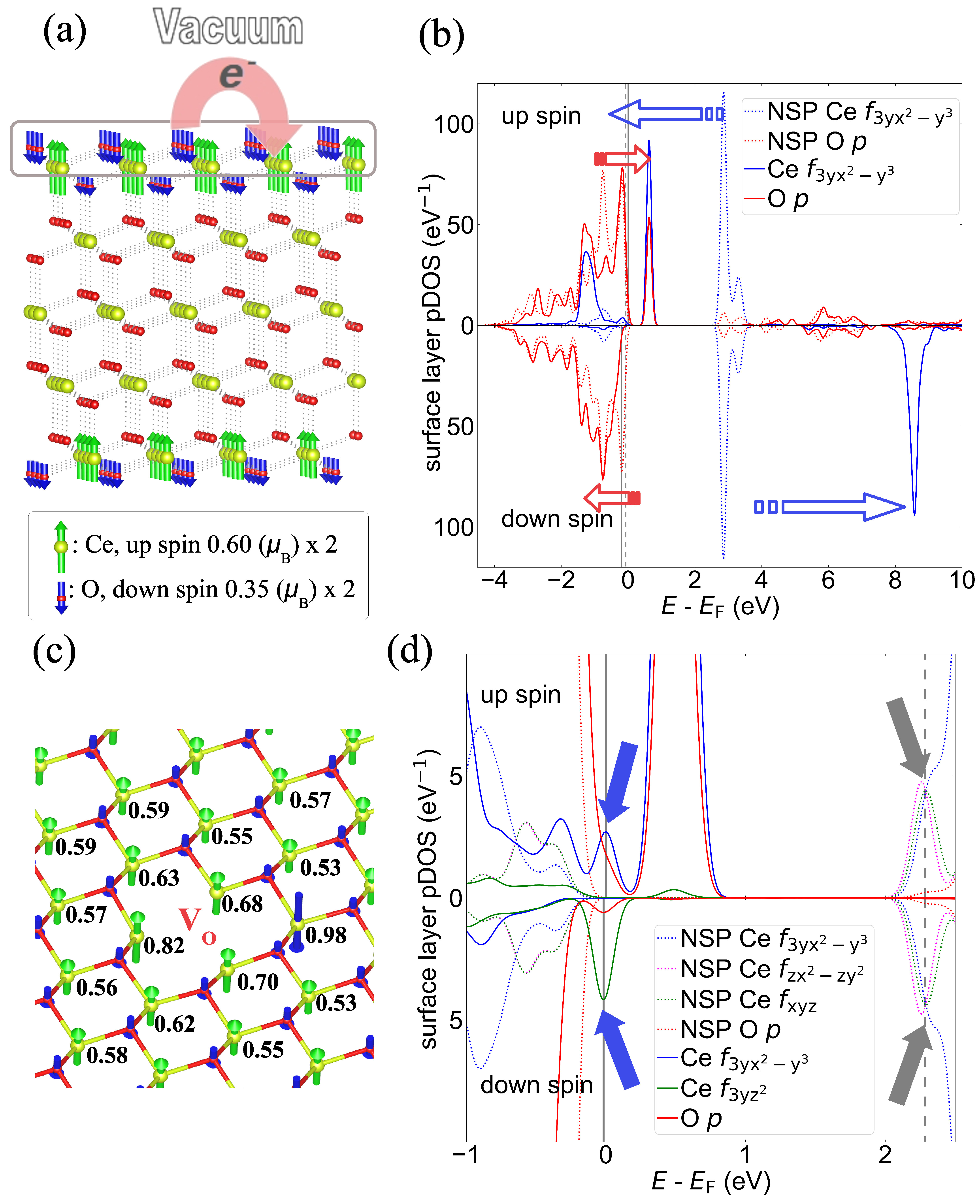}
\caption{\label{Fig_dos} Calculated surface spin polarization and surface oxygen vacancy. (a) Optimized CeO$_2$-(111) surface. Green (blue) vector denotes calculated Mulliken up (down) spin population multiplied by scale factor 2.0 for visualizing the moment (Note that the same scale factor is used for all figures in this work). For clarity, only the spin population larger than 0.20 $\mu_{\textrm{B}}$ is shown here. The lower inset shows the scale of vectors. (b) Calculated surface layer projected density of state. The density of state of upper oxygen and cerium layers marked by grey box in (a) are shown. The blue (red) arrows denote the relative shifts of Ce-$f$ (O-$p$) states between SP and NSP simulations. (c) Optimized surface geometry with oxygen vacancy. (d) Calculated surface layer projected density of state for the oxygen vacancy configuration. The blue (gray) arrows point to defect states obtained from SP (NSP) simulation. In (c) and (d), the zero of energy set to chemical potential of up spin from spin polarized simulation, and relative chemical potentials are marked by solid (dashed) vertical lines for SP (NSP) simulation.}
\end{figure}

\subsection{Surface spin polarization}
We first performed the geometry optimization with initial magnetic moment on all cerium atoms to include surface spin polarization explicitly. Fig. \ref{Fig_dos}a shows the optimized CeO$_{2}$-(111) surface where spins are polarized. The average Mulliken spin moment per atom is 0.60 (-0.35) $\mu_{\textrm{B}}$ on the uppermost cerium (oxygen) layer. This structure contains spin polarization on both top and bottom O-Ce-O surface layers. The spin polarization in the bottom layer (with frozen atoms) is weaker than that in the upper layer (with free atoms), indicating a Jahn-Teller effect ~\cite{guiling_jmmm2018,pentcheva_prl2005,yasemin_jssc2013,zywietz_prb2000}. Fig. \ref{Fig_dos}b shows the projected density of states (DOS) of surface layers. For a comparison, the DOS without spin polarization is also shown here. The spin polarized DOS near the Fermi energy is comprised of appreciable hybridization between Ce-$f$ and O-$p$ orbitals. The occupied (unoccupied) up spin states derived by bonding (antibonding) of Ce-${f}$ and O-${p}$ characters form the narrow bands. The exchange splitting is explained by comprehensive comparison of spin polarized (SP) and non-spin polarized (NSP) DOS. Compared to the NSP system, the Ce-${f}$ states were split into the unoccupied down spin states and occupied up spin state (see the blue arrows in Fig. \ref{Fig_dos}b). In consequence, the net up spin moment is populated on surface cerium atoms. On the other hand, the original O-${p}$ states (from NSP calculation) were pushed up into unoccupied state in up spin and pushed down into lower energy states in down spin (see the red arrows in Fig. \ref{Fig_dos}b), resulting a population of the net down spin moment on surface oxygen atoms.

Compared to the NSP calculation, the inclusion of spin polarization lowers the total energy by 7.83 eV. The energy drop is derived by just outermost two O-Ce-O layers (surface layers) where spins are polarized, since there is no sizeable spin polarization on other layers. Accordingly, the energy gain corresponds to 3.92 eV per surface layer or 0.11 eV per O-Ce-O in the surface layer (there are 36 O-Ce-O units in the surface layer). One can confirm that this estimation is appropriate based on constant total energy difference between SP and NSP systems with respect to number of layers, as shown in Table.~\ref{table_layers}. This significant energy difference indicates the existence of a strong spin polarization on CeO$_{2}$-(111) surface. We did not find any antiferromagnetic configuration in this system. Therefore, we conclude that the CeO$_{2}$-(111) surface is ferromagnetic.

While the room temperature magnetism in CeO$_{2}$ has been studied extensively, its origin remains unclear~\cite{karl_physrep2018}. Here, we propose a new theoretical description of the room temperature ferromagnetism in undoped ceria nanostructure~\cite{anwar_ml2011,mingjie_apl2009,novica_nano2012,sumalin_eml2015,sundaresan_nanotoday2009,yinglin_jpcm2008,zhang_nano2010} and thin films~\cite{fernandes_essl2011,fernandes_jpcm2010,fernandes_prb2009,gao_lang2008,nicholas_rsc2016}. It was suggested that the ferromagnetism is confined to the ceria nanostructure surface where oxygen vacancy may play an important role~\cite{sundaresan_nanotoday2009,sumalin_eml2015,fernandes_essl2011}. On the contrary, Li \textit{et al.} \cite{mingjie_apl2009} and Liu \textit{et al.} \cite{yinglin_jpcm2008} proposed that the ferromagnetism does not relate to the surface oxygen vacancy but to the surface Ce$^{3+}$/Ce$^{4+}$ pairs. A number of DFT studies have attributed the ferromagnetism to native defects such as oxygen vacancy. The defect formation causes spin polarization of ${f}$ electrons for Ce ions near the vacancy, resulting in a nonzero net magnetic moment on the ceria surface~\cite{fernandes_prb2009,ge_apl2008,hachimi,keating_jpcc2011,ribeiro_prb2017,xiaoping_jpcc2016,xiaoping_prb2009}. Nevertheless, based on our results, the ferromagnetism may emerge in a bare surface without any defect or impurity. In general, it was perceived that ions on the surface with lower coordination numbers can cause an unusual spin population ~\cite{renan_ass2018,ameerul_ass2018,munoz_jem2010}. In our case, vacuum acts as a virtual doping on the surface: hole doping for oxygen layer lowering the chemical potential of surface oxygen and electron doping for cerium layer. Upon significant spin exchange interaction on the surface, the virtual doping leads to a charge transfer from oxygen layer to cerium layer, resulting in Ce$^{4+} \rightarrow$ Ce$^{3+}$ transition and partially filled O-${p}$ on the surface layer (see Figure \ref{Fig_dos}a-b). In consequence, the spin on ions couples ferromagnetically, which gains significant energy comparing to the NSP system. In addition, we emphasize that the charge transfer is fully triggered by the surface effect. As shown in Table~\ref{table_layers}, increasing the number of layers reduces the surface/bulk ratio, resulting in an increased chemical potential which approaches to that of the bulk, while other quantities show no sizeable variation.

\begin{table}[ht]
\begin{center}
\begin{threeparttable}    
\caption{The slab size effect on surface spin polarization. The O1 (O2) denotes oxygen atoms on the top (bottom) layer within the surface O-Ce-O layer. The calculated quantities by surface spin polarized (SP) and non-spin polarized (NSP) simulations are presented. $E_{\textrm{surf}}$ is the calculated surface energy. $E_{\textrm{F}}^{\textrm{bulk}}$ and $E_{\textrm{F}}^{\textrm{slab}}$ are the Fermi energies for ceria bulk and slab, respectively. $E_{\textrm{SP}}$ and $E_{\textrm{NSP}}$ are total energies of SP and NSP systems, respectively. \{O-Ce-O\}$_{\textrm{surf}}$ denotes O-Ce-O in the surface layer. 
}\label{table_layers}
    \begin{tabular}{ccccccc}
      \toprule
      \hline\hline
      \multicolumn{2}{c}{} & \multicolumn {5}{c}{~~~~Number of O-Ce-O layers~~~~} \\\midrule
      \cline{3-7}
      \multicolumn{2}{c}{} & 3 & 5 & 7 & 9 & 11\\
      \hline
      \multirow{3}{*}{ \makecell{~~~~Spin population~~~~\\in surface layer\\ ($\mu_{\textrm{B}}$/atom) } } & Ce & 0.55 & 0.60 & 0.58 & 0.56 & 0.56 \\
      & O1 & -0.32 & -0.35 & -0.34 & -0.33 & -0.33 \\
      & O2 & -0.19 & -0.20 & -0.20 & -0.20 & -0.20 \\
      \hline

      \multirow{2}{*}{$E_{\textrm{surf}}$ (J/$\textrm{m}^{2}$)\footnote{a} } & SP & 0.94 & 0.94 & 0.95 & 0.95 & 0.96 \\
      & NSP & 1.07 & 1.08 & 1.08 & 1.08 & 1.08 \\
      \hline
      \multirow{2}{*}{$E_{\textrm{F}}^{\textrm{bulk}}$ - $E_{\textrm{F}}^{\textrm{slab}}$ (eV)} & SP & 8.34 & 6.86 & 5.59 & 4.70 & 4.04 \\
      & NSP & 8.44 & 6.93 & 5.64 & 4.77 & 4.12 \\   
      \hline
      \multicolumn{2}{c}{ \makecell{$E_{\textrm{SP}}$-$E_{\textrm{NSP}}$ \\ (eV/\{O-Ce-O\}$_{\textrm{surf}}$ )} } & -0.11 & -0.11 & -0.10 & -0.10 & -0.10\\ 
      \hline\hline
    \end{tabular}
        \begin{tablenotes}
        \item[a] Experimental surface energy is 1.2 $\pm$ 0.2 J/$\textrm{m}^{2}$ ~\cite{shmuel_jacs2011}.
    \end{tablenotes}
    \end{threeparttable}
\end{center}
 
\end{table}

\begin{figure*}
\centering
\includegraphics[width=0.99 \textwidth]{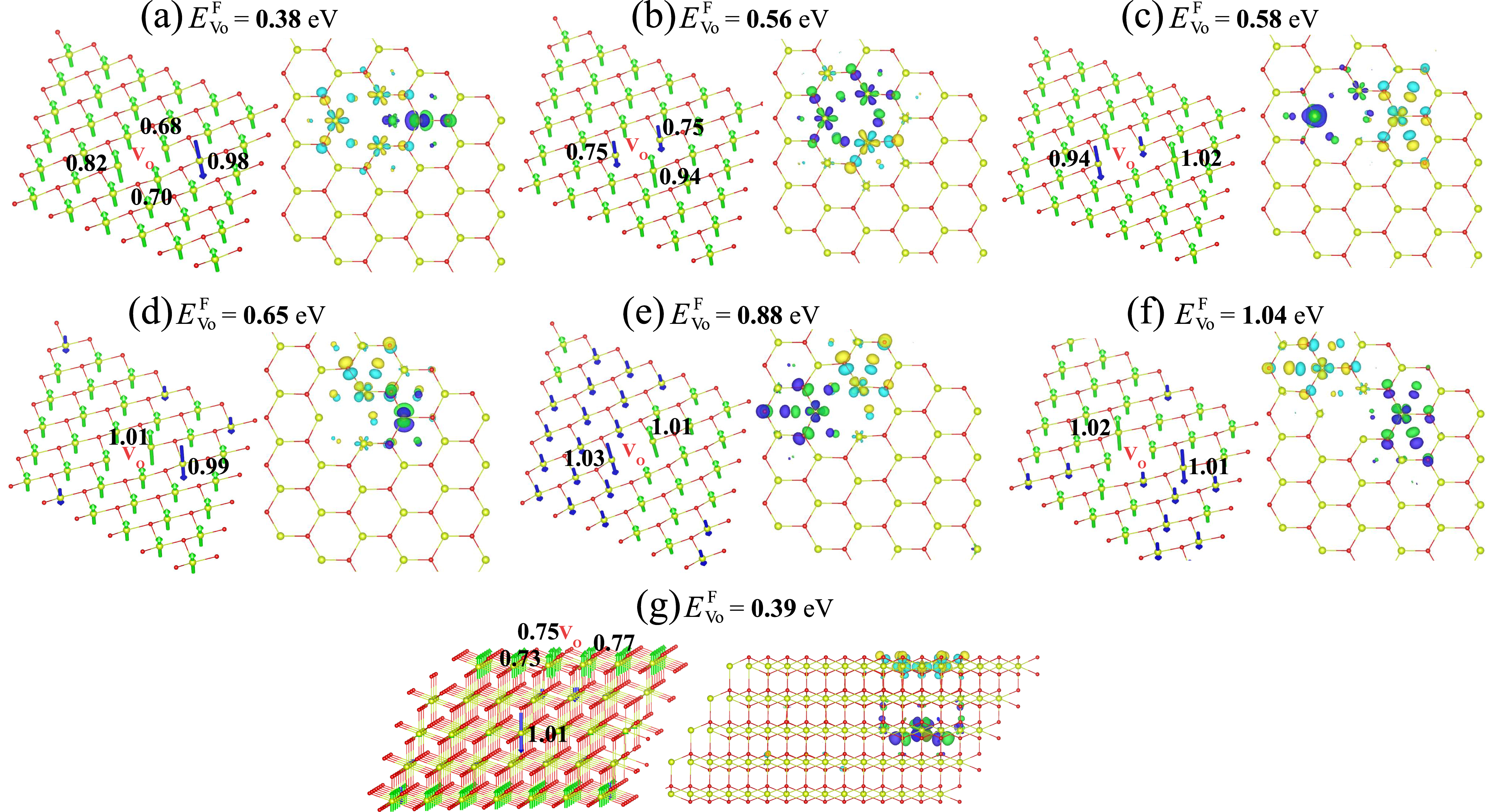}
\caption{\label{Fig_ovlist} Low-energy oxygen vacancy structures. In the sub-figures, the left panel shows the calculated Mulliken up (down) spin population ($\mu_{\textrm{B}}$) denoted by Green (blue) vector. The right panel shows wavefunction isosurface ($0.05 e/\textrm{\AA}^{3}$) of the localized defect states. The yellow/cyan (green/blue) correspond to positive/negative wavefunction of spin up (down)  defect state. (a) the most stable configuration, one spin up excess electron at three nearest neighbor Ce atoms and the other spin down excess electron at next nearest neighbor Ce atom, (b) one spin down excess electron at nearest neighbor two Ce atoms and the other spin up excess electron at nearest neighbor Ce atom, (c) one spin down excess electron at the nearest neighbor Ce atom and the other spin up excess electron at next nearest neighbor Ce atom, (d) one spin up excess electron at the nearest neighbor Ce atom and the other spin down excess electron at next nearest neighbor Ce atom, (e) two excess electrons at nearest neighbor Ce atoms, (f) two excess electrons at next nearest neighbor Ce atoms, and (g) one spin up excess electron at three nearest neighbor Ce atoms and the other spin down excess electron at Ce atom in the middle layer.}
\end{figure*}

On account of the robust surface spin coupling, we investigated the impact of surface spin polarization on surface oxygen vacancy formation. Figure \ref{Fig_dos}c-d show the optimized structure, spin population and DOS of surface layer with oxygen vacancy. In the SP simulation, the two excess electrons produced by the oxygen vacancy are populated on neighboring cerium atoms forming localized polaronic states. One electron was captured by three nearest neighboring Ce atoms, with the spin moments of 0.82, 0.68 and 0.70 $\mu_{\textrm{B}}$, respectively. The other electron was localized on the next nearest neighboring Ce atom with 0.98 $\mu_{\textrm{B}}$ down spin moment having antiferromagnetic coupling with the surrounding Ce atoms (see Fig.~\ref{Fig_ovlist}a). The latter is associated with the increased local spin moment by Hund's rule within our singlet simulation. The oxygen vacancy defect states in the SP system are shallow, whereas the same defect states are below conduction band minimum in the NSP system. With this configuration, we found the lowest surface oxygen vacancy formation energy is 0.38 eV in the surface spin polarized system. In contrast, the NSP simulations show higher oxygen vacancy formation energy values: 4.06 eV ($U$ = 7 eV) and 3.04 eV ($U$ = 4 eV), as compared to several previous reports ranged from 2.13 to 3.20 eV \cite{song_surfcesi2013, hui_prb2009, zhang_nano2010, veronica_prl2009}.  Note that two electrons are almost equally populated on two nearest neighbor Ce atoms and one next nearest neighbor Ce in our NSP simulation. In the SP system, the less ionized oxygen ions bond weakly with cerium ions, resulting in lower cost of vacancy formation in comparison to NSP system which does not involve charge transfer process. Therefore, our results indicate that surface spin polarization can lower the oxygen vacancy formation energy, and enhance the oxide reduciblility of the CeO$_{2}$-(111) surface. 
 
\begin{figure}[ht]
\centering
\includegraphics[width=0.45 \textwidth]{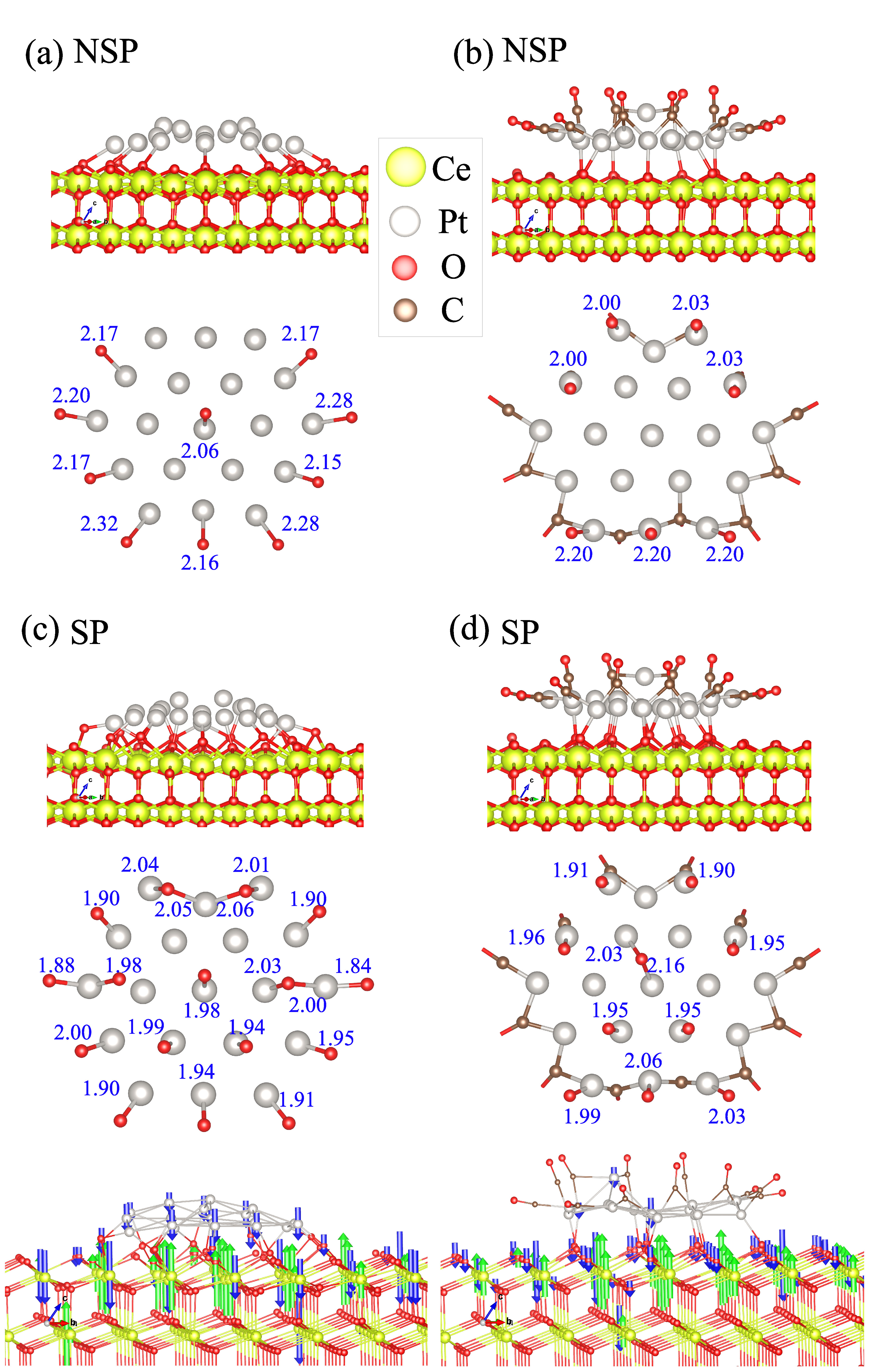}
\caption{\label{Fig_SL} Optimized Pt single layer on ceria. (a) Pt$_{19}$ single layer on a non-spin polarized CeO$_2$-(111) surface. (b) Pt$_{19}$[CO]$_{12}$ on a non-spin polarized CeO$_2$-(111) surface. (c) Pt$_{19}$ single layer on a spin polarized CeO$_2$-(111) surface. (d) Pt$_{19}$[CO]$_{12}$ on a spin polarized CeO$_2$-(111) surface. The side view of the structures are shown in the top panel. The calculated bond lengths ($\textrm{\AA}$) between Pt and surface oxygen are shown in the middle panel.}
\end{figure}

In addition to the most stable oxygen vacancy defect state, we found several other low-energy spin configurations. In Fig.~\ref{Fig_ovlist}c, local spin moments on two Ce atoms are exchanged in comparison to the most stable configuration in Fig.~\ref{Fig_ovlist}a. This vacancy configuration gives rise to down spin flipping on another nearest neighbor Ce atom and cause an increase in formation energy of 0.2 eV from the most stable oxygen vacancy. In the configuration of Fig.~\ref{Fig_ovlist}b, one electron was equally distributed on two nearest neighbor Ce atoms which have down spin moment of 0.75 $\mu_{\textrm{B}}$ opposite to the majority up surface spin. This configuration leads to increased formation energy of 0.18 eV in comparison to the most stable oxygen vacancy. These indicate that surface spins favor ferromagnetic coupling with excess electrons. In Figure~\ref{Fig_ovlist}d-f, we found very localized defect states with a Mulliken population close to 1.0 $\mu_{\textrm{B}}$ on two Ce atoms. However, that localization does not only weaken ferromagnetic coupling by producing opposite spin moments on the surface, but also increases the repulse Coulomb interaction between excess electrons, resulting in a higher formation energy. In particular, the configuration of Fig.~\ref{Fig_ovlist}f, where two excess electrons are localized on next nearest neighbor Ce atoms, was reported as the most stable oxygen vacancy state by other DFT studies~\cite{song_surfcesi2013,hui_prb2009,zhong_prm2018}. However, the configuration is no longer the ground state by considering the spin effect. As shown in Fig.~\ref{Fig_ovlist}g, we also found outstanding oxygen vacancy configuration to be energetically degenerated with the most stable oxygen vacancy configuration. One defect state located on three nearest neighbor Ce atoms increasing spin moment ~0.75 $\mu_{\textrm{B}}$ around oxygen vacancy in the ferromagnetic coupling, whereas the other excess electron is moved to the farthest site from surface, which has no spin coupling. In comparison to total spin population on oxygen layer of the most stable configuration, this configuration further increased total spin moment of 0.58 (0.39) $\mu_{\textrm{B}}$ on the uppermost (middle) two oxygen layers. This ascertains additional charge transfer to localize the excess electrons. These configurations suggest that even in the presence of oxygen vacancies,  there is still surface spin polarization, and it can even be enhanced for some vacancy configurations as shown in Fig.~\ref{Fig_ovlist}.     

\begin{figure*}[!htb]
\centering
\includegraphics[width=0.98 \textwidth]{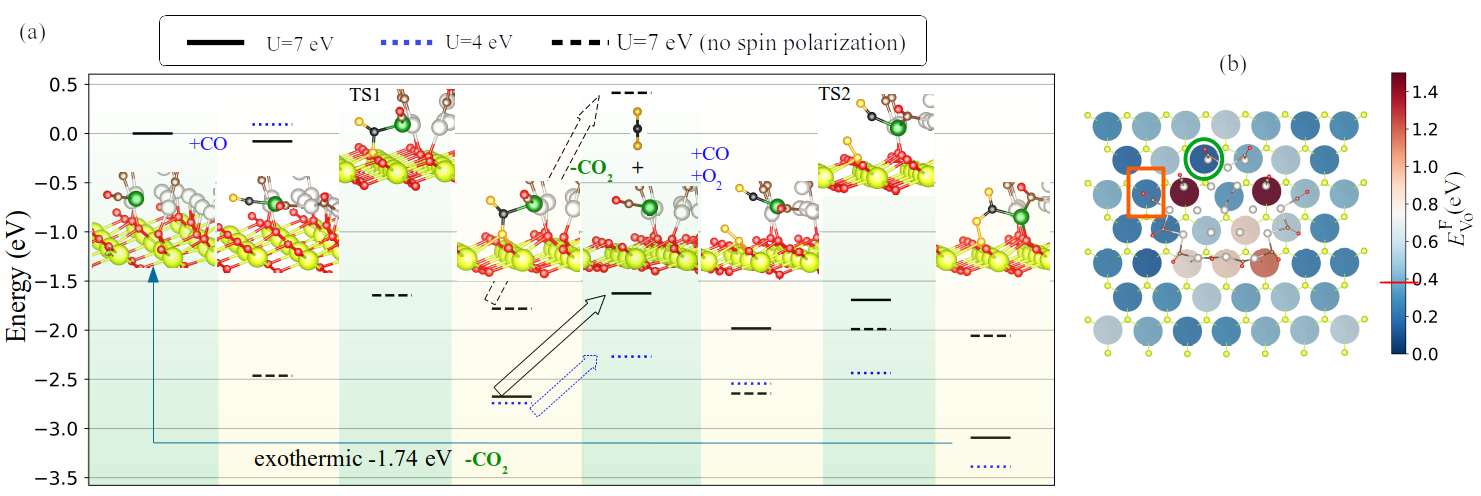}
\caption{\label{Fig_MvK} (a) The calculated energy profiles for proposed reaction pathway for CO oxidation on Pt$_{19}$[CO]$_{11}$ with the ceria support. The black solid (blue dotted) lines show the calculated energy profiles by using $U$ = 7.0 (4.0) eV from the spin polarized simulations. The black dashed lines show the calculated energy profiles by using $U$ = 7.0 eV from the non-spin polarized simulations. TS1 and TS2 denote two transition states. CO is absorbed on green colored Pt atom. C (O) atoms involved in the reaction were colored by black (yellow). (b) Calculated surface oxygen vacancy formation energies within SP simulations. Filled circles represent the formation energies at corresponding sites. The surface oxygen which react with CO was marked by the red box. The surface oxygen with exceptional low vacancy formation energy was marked by the green circle. The energy level of oxygen vacancy formation for the most stable configuration on bare ceria surface is marked by red line on the energy color bar.}
\end{figure*}

\subsection{Pt on the ceria support with CO}

To date, surface spin polarization has not been taken into account for nano-structured heterogeneous catalysts. Hence we investigated the influence of surface spin polarization on the interaction between Pt nanoparitcle and ceria. As shown in Fig.~\ref{Fig_SL}a, the optimized Pt$_{19}$ SL on CeO$_{2}$-(111) consisted of platinum and surface oxygen (O$_{\textrm{surf}}$) bonds mostly on perimeter of the SL in NSP simulation. On the other hand, the SP simulation (see Fig.~\ref{Fig_SL}c) generates a configuration with more Pt-O$_{\textrm{surf}}$ bonds. In addition, the averaged bond length is also shorter than that of the NSP configuration. This indicates that the interaction between ceria surface and Pt$_{19}$ SL is significantly enhanced by the surface spin polarization. We attribute the strong Pt-O$_{\textrm{surf}}$ bonds to electron depletion oxygen layer by the charge transfer on the spin polarized surface, where Pt atoms are more prone to oxidation. The enhanced interfacial interaction induces spin moments on Pt and Ce atoms and coupling between them. The surface oxygens were pulled out from ceria surface by this interaction. As a result, the donation of Pt valance electrons to surface oxygens increases the unpaired spin density around the Pt site. The pronounced on-site spin populations on Pt atoms are denoted in Fig. \ref{Fig_SL}c. The interfacial interaction also affects the electronic configuration of Ce atoms on the surface layer. The less oxidized Ce atoms has a large unpaired spin population which couples with the polarized Pt spins through surface oxygen. The majority of magnetic coupling is antiferromagnetic, resulting in a zero net spin moment on the intermediate surface oxygen.  

Our simulation shows that adsorption energy of CO on the spin polarized CeO$_{2}$-(111) surface is -0.31 eV (see Fig. S1). The result agrees with previous observations that CO on CeO$_{2}$-(111) surface is either unstable or has weak interaction with the surface~\cite{fendy_pccp2012,huang_jpcc2008,michael_jpcc2006} (see Section B in the SI). In terms of CO's adsorption energy on the ceria-supported Pt nanoparticle, our calculation reveals a wide distribution from -3.14 to -1.02 eV, depending on the choice of adsorption sites (see Fig. S2). These results indicate that CO molecules are strongly adsorbed on Pt nanoparicles~\cite{yubing_ccc2020}. In this work, we constructed ceria-supported Pt$_{19}$ SL with 12 CO molecules which were initially located at the bridging sites of the perimeter of Pt$_{19}$. Optimized geometries of Pt$_{19}$[CO]$_{12}$ with/without surface spin polarization are shown in Fig. \ref{Fig_SL}b and d. We again observed that surface spin polarization enhances the interaction between Pt$_{19}$[CO]$_{12}$ and ceria, i.e., more bonds and shorter bond lengths. 



Fig. \ref{Fig_SL} shows that the CO adsorption on Pt$_{19}$ breaks a few Pt-O$_{\textrm{surf}}$ bonds on the perimeter of Pt$_{19}$ in both NSP and SP calculations. Thus, the adsorption weakens the interfacial interaction. The weakening can be understood from Mulliken spin population analysis in the SP system. The strong CO adsorption on Pt$_{19}$, due to the $\pi$-back bonding between Pt and CO~\cite{suparna_crssms2015}, can reduce the donation of Pt's valence electron to surface oxygen. As a result, the pairing electrons on Pt atoms decreases the net spin moment on the Pt$_{19}$ (see Figure \ref{Fig_SL}c-d). 

\subsection{Reaction pathway in CO oxidation}

\begin{table}[]
\caption{The list of activation barriers in energy profiles for proposed reaction pathway. The major activation barriers are hilighted in bold.}\label{table_energy}
\begin{tabular}{lcccc}
\hline\hline
Calculation Type& Parameter & \multicolumn{3}{c}{Reaction Barrier (eV)} \\
                     & $U$ (eV)    & ~~~TS1~~~   & ~~~-CO$_2$~~~    & ~~~TS2~~~      \\\hline
Spin Polarization         & 7.0         & 0       & \textbf{1.05}      & 0.29     \\
Spin Polarization         & 4.0         & 0       & \textbf{0.47}      & 0.11     \\
Non-spin Polarization     & 7.0         & 0.82      & \textbf{2.19}      & 0.65   \\
\hline\hline
\end{tabular}
\end{table}
We found that surface spin polarization may play an important role in the catalytic activity. On a spin polarized surface, the electrophilic oxygens in the electron deficient condition are more reactive with other species in comparison to the NSP system. In addition, new species formed by the reaction are easily removable owing to the high reducibility on the spin polarized surface. The impact of spin polarization on surface catalytic activity is illustrated in Fig. \ref{Fig_MvK}a. We started with a CO desorpted Pt$_{19}$[CO]$_{11}$ with ceria support in CO rich conditions. The Pt$_{19}$[CO]$_{11}$ can adsorb one more CO molecule to form Pt$_{19}$[CO]$_{12}$, with a shallow adsorption energy of -0.08 eV. The adsorbed CO reacts with O$_{\textrm{surf}}$ to form CO$_{2}$, gaining an energy of 2.60 eV. While there is 
no activation barrier for the reaction in the SP system, \textbf{}calculated activation barrier is 0.82 eV in the NSP system. Taking off the CO$_{2}$ molecule from the reaction site leaves a surface oxygen vacancy, costing an penalty energy of 1.05 eV for $U$ = 7 eV (0.47 eV for $U$ = 4 eV). This is the main barrier in the proposed reaction pathway\textbf{.} The Pt SL can subsequently adsorb another incoming CO, while the ceria surface can attract O$_2$ at the oxygen vacancy site. The molecular O$_2$ and CO further react to form a new CO$_{2}$. The calculated activation barrier for the reaction are 0.29 ($U$ = 7 in the SP system), 0.11 ($U$ = 4 in SP system), and 0.65 ($U$ = 7 in the NSP system) eV, respectively. Thereafter, the catalytic cycle is completed by desorption of the CO$_{2}$, which is an exothermic process releasing energy of 1.74 eV to recover to Pt$_{19}$[CO]$_{11}$. This proposed reaction pathway is manifested by MvK mechanism~\cite{mars_ces1954}. The simulated reaction pathway using $U$ = 4 eV has the same trend with $U$ = 7 eV, as shown in Fig.~\ref{Fig_MvK}a. However, the NSP system shows different reaction energy profiles from the SP system.  Table.~\ref{table_energy} summarizes the activation barriers for each system. The major activation barriers (0.47-1.05 eV) of the SP system are apparently close to the measured activation energies ranging from 0.37 to 0.77 eV~\cite{yubing_ccc2020,jun_acscata2015,matteo_science2013,rene_angcom2015,vincent2021fluxional}, while the NSP system has a much higher barrier (2.19 eV). The elevated barrier in the NSP system is caused by high formation energy of oxygen vacancy, that hampers the release of CO$_{2}$ from the surface.   

\begin{figure*}[ht]
\centering
\includegraphics[width=0.98 \textwidth]{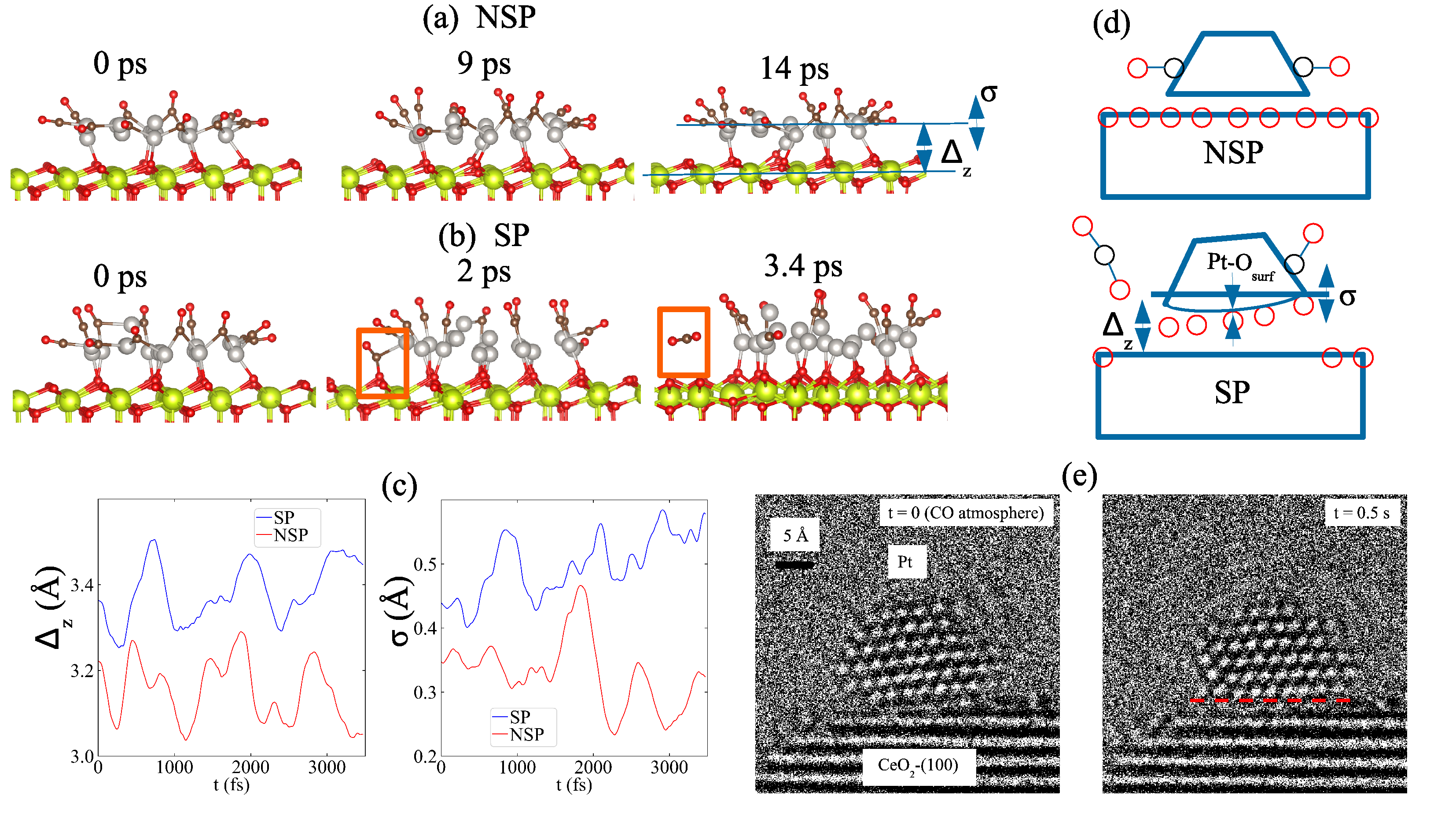}
\caption{\label{Fig_md} (a) The snap shot of NSP AIMD simulation of the ceria-supported Pt$_{19}$ single layer with 12 CO molecules at 0, 9 and 14 ps. (b) The snap shot of SP AIMD simulation of the ceria-supported Pt$_{19}$ single layer with 12 CO molecules at 0, 2 and 3.4 ps. The red box shows that the CO and surface O bind to form CO$_{2}$. (c) Left panel: calculated distance between the average z coordinates of surface Ce atoms and the average z coordinates of Pt atoms, right panel: standard deviation of z coordinates of Pt atoms. (d) Schematic diagram of lattice dynamic in SP and NSP systems. (e) Atomic resolution transmission electron microscope images of fluxional Pt nanoparticle on CeO$_{2}$-(100) surface in a CO atmosphere (7 x 10$^{-4}$ Torr) at room temperature. Pt columns are visible as white dots whereas (100) Miller planes in CeO$_{2}$ appear as white horizonal lines. The two images are from the same nanoparticle with right-hand image recorded ~0.5s after the left-hand image. The Pt nanoparticles undergoes 7.5\textdegree ~clockwise rotation resulting in Pt (111) becoming approximately parallel to (100) CeO$_{2}$ plane (see red dotted guide to eye on right-hand panel). The Pt atomic plane near the CeO$_{2}$ is not straight and curves up on left-hand side and also the atomic columns on the right are slightly blurred suggesting a high degree of instability in their position. Other changes in the Pt surface structure are apparent due to the strong interaction with CO.}
\end{figure*}

In addition to lowering the activation barrier, the surface spin polarization enables the adsorbed CO on Pt SL to react with O$_{\textrm{surf}}$ to form CO$_{2}$. While this reaction is endothermic in the NSP system, the energy released by the reaction is large in the SP system, owing to high surface oxygen reactivity. Compared to the most stable oxygen vacancy on a bare ceria surface, the surface oxygen (at the reaction site marked by red box in Fig.~\ref{Fig_MvK}b) vacancy formation energy with Pt$_{19}$[CO]$_{12}$ is lowered by 0.15 eV, thus significantly boosting the reaction with surface oxygen. In Fig.~\ref{Fig_MvK}b, we show the calculated vacancy formation energies for all surface oxygen sites. 
The interaction between Pt$_{19}$[CO]$_{12}$ and CeO$_{2}$ generally lowers the vacancy formation energies for surface oxygen at exterior sites of the contact zone. However, within the contact zone, the majority of surface oxygens strongly bond with Pt ions (see Fig.~\ref{Fig_SL}d). The vacancy formation energies are relatively high for these oxygen, except the oxygen which bonds with Pt-CO (marked by green circle in Fig.~\ref{Fig_md}b). The O$_{\textrm{surf}}$-Pt-CO at the exceptional site may migrate by forming dynamic low-coordinated atoms, which has been proposed as the prime cause of dynamic structure of ceria supported gold nanoparticles upon exposing to CO and oxygen gases~\cite{he2018size}. The impact of surface spin polarization on structural dynamics will be discussed in the following section. 

\subsection{Structural dynamics}
The dynamic lattice reconfiguration of nanoparticles upon exposing to oxidizing and reducing gases (e.g. CO and O$_{2}$) has been intensively studied by Transmission Electron Microscopy (TEM) in the recent years~\cite{poul_science2002,yasutaka_angechem2008,vendelbo_nmat2014,andreas_jcata2015,andreas_angchem2017,he2018size,lawrence2018oxygen,lawrence2021atomic,li2021dynamic,miller2021linking,tao2016atomic,vincent2020atomic,vincent2021fluxional}. These studies imply that the surface and perimeter of nanoparticle strongly interact with CO and O$_{2}$ gases. To explore the potential influence of surface spin polarization on the dynamic lattice reconfiguration, we performed two independent AIMD simulations of ceria-supported Pt nanoparticles for 20 ps with and without the inclusion of spin polarization. Due to the limitation of computational resources, we used only three layers of O-Ce-O. Therefore, the CeO$_2$ support is highly reducible, resulting in a unrealistically fast CO oxidation even at 300 K in our AIMD simulation. 

Despite the artefact due to the choice of a thin CeO$_2$ slab, our simulations still reveal distinct physical pictures due to the inclusion of spin polarization. In the NSP AIMD simulation (Fig.~\ref{Fig_md}a), the whole system does not show any obvious structural change for 20 ps. On the other hand, we observed rapid CO$_2$ formation at around 3 ps (Fig.~\ref{Fig_md}b) when the spin polarization is turned on. Fig.~\ref{Fig_md}c plots the calculated distance between the averaged $z$ coordinates of surface Ce/Pt atoms, as well as the standard deviation to evaluate dispersion of $z$ coordinates of Pt atoms during the simulation. Clearly, the Pt atoms in the SP system are more distant from surface Ce atoms with a larger fluctuation, in comparison to the NSP system. Accordingly, we identify four different features of the SP system (in schematic Fig.~\ref{Fig_md}d). First, strong bonds between Pt and surface oxygen atoms are found in the SP system. This is shown by more bonds and shorter bond lengths because oxygen has moved towards the Pt layer as seen in Fig.~\ref{Fig_SL}. Second, the large distances between surface Ce and Pt atoms are present consistently in the SP AIMD simulations. Highly reducible surface oxygen in the SP surface bond strongly with Pt atoms, resulting in an elevation of surface oxygen and Pt atoms from CeO$_{2}$ surface. This shifting is also manifested in the optimized structures in Fig.~\ref{Fig_SL}. Third, substantial fluctuations of Pt atoms are displayed in the SP AIMD simulations, indicating that the elevated Pt nanoparticle from CeO$_{2}$ surface are flexible. Fourth, asymmetric CO oxidation was raised by irregular structure of Pt and nonuniform surface O vacancy formation energies as shown in Fig.~\ref{Fig_MvK}b.

Recently, experimental evidence has been published that shows significant lattice reconfiguration on Pt nanoparticles supported on CeO$_{2}$ on exposure to CO and other gases~\cite{li2021dynamic,vincent2020atomic,vincent2021fluxional}
While the timescales for the computation and experimental datasets are very different, precluding a detailed quantitative structural comparison, there is qualitative agreement in the trends in structural dynamics from both theory and experiment. Fig.~\ref{Fig_md}e is in situ electron microscopy data  showing dynamic structural change taking place in a CO atmosphere (for experimental details see reference \cite{vincent2020atomic}). The Pt nanoparticles undergoes complex lattice changes such as clockwise rotation, uneven Pt atomic plane near the CeO$_{2}$, instability its atomic position, and apparent change on the Pt surface.  There are also continuing changes in the interface structure due to the constant creation and annihilation of oxygen vacancies at  perimeter sites.  This occurs due to the weakening of the interfacial bonds in the presence of CO and the experimental observation can be understood in terms of the schematic diagram of Fig.~\ref{Fig_md}d. There are many other fluxional observations such as dynamic changes in cation positions due to breaking and formation of chemical bonds at the perimeter sites which are described in greater detail in the author’s paper~\cite{vincent2021fluxional} . These complex dynamics may be described by concerted effect from aforementioned four features, which are pronounced in our SP AIMD simulation. Thus, we suggest that surface spin polarization may play a deterministic role in promoting the dynamic lattice reconfiguration of ceria-supported Pt nanoparticles.

\section{Conclusions}
In summary, we report the robust surface spin polarization on the CeO$_{2}$-(111) surface due to charge transfer from surface oxygen to cerium. The surface spin polarization appears to be essential to describe the existing important observations such as ferromagnetism in undoped CeO$_{2}$ nanostructures and thin films, high reducibility of ceria support, and low CO oxidation reaction barrier. In addition, the surface spin polarization enhances the bonding between platinum and surface oxygen. However, the CO adsorption on the perimeter of platinum single layer is inclined to weaken the interfacial interaction. We expect the presence of vigorous surface spin polarization at ambient temperature will be useful to understand the interfacial interaction and guide the design high-performance heterogeneous catalyst.

\section*{Acknowledgments}
The computing resources are provided by XSEDE (TG-DMR180040). The authors gratefully acknowledge financial support from the National Science Foundation (NSF). BK and QZ are supported by NSF-OAC award 1940272. NSF-CBET Award 1604971 and NSF-OAC Award 1940263 supported JLV and PAC, who acquired and processed the experimental data and participated in extensive discussion of the calculations.  The authors thank Arizona State University’s John M. Cowley Center for High Resolution Electron Microscopy for microscope access and use.




\bibliography{ref}

\end{document}